\documentclass[floats]{revtex4}

\usepackage{graphicx}

\begin{document}

\title {$1/T_1$ nuclear relaxation  time of  $\kappa-(BEDT-TTF)_ 2 Cu [N(CN)_2] Cl$ :  effects of magnetic frustration}

\author{I. J. Hamad$^1$,  A. E. Trumper$^1$ , P. Wzietek$^2$, S. Lefebvre$^2$, and L. O. Manuel$^1$}
\affiliation {
$^{(1)}$Instituto de F\'{\i}sica Rosario (CONICET) and
Universidad Nacional de Rosario, Boulevard 27 de Febrero 210 bis, (2000) Rosario, Argentina.\\
$^{(2)}$ Laboratoire de Physique des Solides (CNRS, U.R.A.2), Universit\'e de Paris-sud, B\^atiment 510, 91405 Orsay, France. }

\vspace{4in }
\date{\today}

\begin{abstract}
We study the role played by the magnetic frustration  in the antiferromagnetic phase of the organic salt $\kappa-(BEDT-TTF)_ 2 Cu [N(CN)_2] Cl$. Using the spatially anisotropic triangular Heisenberg model we analyze  previous and new performed NMR experiments. We compute the  $1/T_1$ relaxation time by means of the modified spin wave theory. The strong suppression of the nuclear relaxation time observed experimentally   under varying pressure and magnetic field
is qualitatively well reproduced by the model. Our results suggest the existence of a close relation between the effects of pressure and magnetic frustration. 
\end{abstract}
\maketitle

\section{Introduction}
\begin{figure}[ht]
\includegraphics[width=8cm]{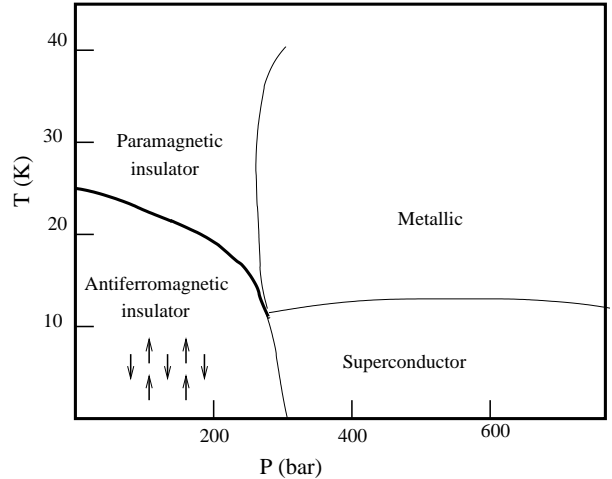}
\caption{Schematic temperature vs pressure phase diagram of $\kappa - Cl$ obtained in ref. \cite{lefebvre}. The thick line represents the magnetic transition of interest for the present study.}
\label{diagrama}
\end{figure}

The interplay between frustration and strong correlation in electronic systems has become a central issue
in condensed matter theory. Among the compounds  which manifest this interplay are the quasi-bidimensional 
organic salts $(BEDT-TTF)_2 X$ \cite{mckenzie} and the cobaltate compounds $Na_xCoO_2$\cite{cava}. 
In particular, the $\kappa$ family of the  organic salts displays a molecular arrangement  characterized by 
a strong dimerization of BEDT-TTF molecules in anisotropic triangular layers.
The presence of the monovalent anion $X$ introduces a hole into each dimer rendering the antibonding
molecular orbital of the dimer half filled. Recently,  the phase diagram for $X=Cu[N(CN)_2]Cl$ (hereafter, $\kappa-Cl$) has been obtained\cite{lefebvre} with
paramagnetic insulating (PI), antiferromagnetic insulating (AF), superconducting (SC), and metallic (M) phases (Fig. \ref{diagrama}). The boundary separating the PI from the metallic phase is a first order Mott transition with a critical endpoint at around $40 ^{\circ}K$ and $280$ bar \cite{Limelette}. On the other hand, within a range of pressure of $200-400$ bar, there is a coexistence region of AF and SC phases  which is not shown in Fig. \ref{diagrama}.    
Regarding the insulating phases, as temperature is decreased, there is a transition from PI to an antiferromagnetic phase with a commensurate magnetic order, while the effect of pressure is to decrease the
N\'eel temperature.   
It is worth noting that the paramagnetic insulating phase is always in between the antiferromagnetic and the metallic phases. This  absence
of boundary confirms
the lack of itinerant AF in this organic salt. For this reason, a description
of the PI-AF transition in terms of interacting localized
spins on dimers lying on a triangular lattice deserves further investigation.
It can be hypothesized that the decrease of the N\'eel temperature with pressure may originate from the pressure-induced increase of the magnetic frustration.
While the interplay between pressure, electronic correlation, and frustration  has not yet been elucidated, due to the complex structure of the $\kappa$ compounds, some results in the literature suggest that strong correlation and pressure-induced frustration is at play in the $\kappa$ compounds. For example, it is generally believed that the application of hydrostatic or chemical pressure (i.e., change of anion $X$ composition) reduces the ratio $U/W$, where $U$ is the effective dimer Coulomb repulsion and $W$ is the bandwith (the pressure enhances the inter-dimer integral transfer $t$\cite{campos96}), therefore driving the system through 
a Mott transition\cite{mckenzie}. Very recently, several authors, within the context of Hubbard models,  have proposed a RVB theory of superconductivity for the $\kappa$ organic compounds, and they have emphasized the role of frustration in the transition from the AF to  the metallic phase through a superconducting state. In particular, Powell and McKenzie\cite{powell05} pointed out that the Mott transition can be driven by an increase of frustration even at fixed $U/W$, whereas, Gan {\it et al.}\cite{gan05} proposed 
that the effect of pressure in the phase diagram is to decrease $U/W$ and/or to increase the frustration.

Regarding the insulating phase, the proper effective Hamiltonian
proposed to describe the antiferromagnetic phase of the
$\kappa$ family
is the spatially anisotropic Heisenberg model on a triangular
lattice where each site represents a dimer\cite{mckenzie}, and the exchange interaction is $J=4t^2/U$. The zero temperature
phase diagram of this model has been studied with spin wave theory \cite{trumper}, Schwinger bosons \cite{manuel}, and series expansion \cite{weihong}, and it
shows collinear AF, disordered, incommensurate  and commensurate
spiral phases depending on magnetic frustration.
The AF magnetic phase of the $\kappa-Cl$ compound
 seems to be located at the collinear AF side of the diagram.
Even though the ground state properties of this frustrated microscopic model
have been investigated in the last years, its relevance to describe the insulating phase of the $\kappa$ family
has  been little explored in the literature\cite{McKenzie-condmat}.
 
Here we compare qualitatively the predictions of the aforementioned frustrated model  with our measurements of 
the magnetic field dependence of the $1/T_1$ relaxation time  for  the $\kappa-Cl$ compound. We also compare 
our theoretical results with previous measurements of $1/T_1$  performed under varying pressure\cite{lefebvre}. 
We have solved the model  using  the modified linear spin wave theory  in
 order to avoid both, the
presence of long range order --due to low dimensionality-- and the divergence of the number of magnons at finite temperature. This can be achieved by imposing a zero
magnetization constraint through a Lagrange multiplier \cite{takahashi}. The modified spin wave theory recovers the expected behavior of several thermodynamic properties  for an ample range of
temperature and frustration. The $1/T_1$ nuclear relaxation
time has been computed taking into account the
Raman processes which, by energy
conservation considerations, involve two magnons (simultaneous
creation and destruction of magnons)\cite{pincus}.

 Experimentally, the  location of the $1/T_1$ peak signals the magnetic ordering temperature. The responsible of such magnetic order in this highly anisotropic compound  are the strong correlations or the 2D spin fluctuations present in the organic layers, while the existence of a finite ordering temperature is due to 3D residual interactions. On the other hand, our calculation of the relaxation time is performed in a two dimensional model, and this
fact prevents the existence of an actual phase transition.  However, since our model takes into account the relevant 2D magnetic interactions, the crossover that gives rise to the $1/T_1$ peak  corresponds to what actually happens in the critical region of the real 3D compound.

Our main finding is that the frustrated spin model  reproduces qualitatively well the effects of pressure and magnetic field on the $1/T_1$ relaxation time of the $\kappa-Cl$ compound.
 These results suggest the idea that frustration and  pressure are closely related. Furthermore, the experimental and theoretical results show a strong suppression of the nuclear relaxation time with  increasing magnetic field. This behavior, characteristic of slow spin dynamics, signals the presence of magnetic frustration in the $\kappa$ compound.
 
The article is organized as follows: In Sec. II we develop
the modified spin wave theory for the microscopic model and 
we study the reliability of the approximation by computing
several thermodynamic observables. In Sec. III we analyze the effects of pressure and magnetic field on the $1/T_1$ relaxation time of the AF phase of the $\kappa-Cl$ compound, using the frustrated spin model. The concluding remarks are presented in Sec. IV.

\section{Modified spin wave theory}

\begin{figure}[ht]
\includegraphics[width=5cm]{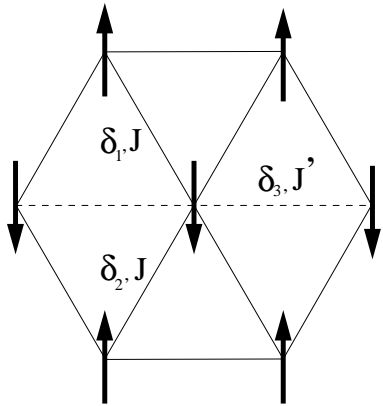}
\caption{N\'eel order characterized by the magnetic wave vector ${\bf{Q_{col}}}
=(0,2\pi/\sqrt{3})$. The doted lines represent the frustrated bonds.}
\label{hexa}
\end{figure}

The spatially anisotropic
Heisenberg model on the triangular lattice is:

\begin{equation}
 {{H}}=\sum_{\bf r,\delta} \;\;J_{\delta}\;\;
 {\bf S}_{{\bf r}}\cdot {\bf S}_{{\bf r}+{\bf \delta}}
 \label{heis}
\end{equation}
where each site represents a half filled dimer,  and  
$J_{\delta}$  are the exchange
antiferromagnetic interactions, $J_{\delta_1}=J_{\delta_2}=J$
and $J_{\delta_3}=J^{\prime}$  (see Fig. \ref{hexa}).

Within the linear spin wave theory the ground state of this model presents two kind of phases\cite{trumper},
the collinear, characterized by the ${\bf Q_{col}}
=(0,2\pi/\sqrt{3})$ wave vector (as shown in Fig. \ref{hexa}), stable in the region
$0\leq J^{\prime}/J\leq0.5$, and the incommensurate spiral,
characterized by ${\bf{Q_{sp}}}=(2Q,0)$, where 
$Q=\cos^{-1}(J/2J^{\prime})$, and stable in the region
 $0.5\leq J^{\prime}/J \leq\infty$. In previous NMR experiments of $\kappa-Cl$  it was observed 
 a splitting of the spectra into a discrete number of lines which can be identified with a commensurate AF order lying in the organic 
 layers\cite{miyagawa}. 
 So, in what follows we
 concentrate our
 attention on the values of frustration interaction that give rise 
 to a collinear structure ${\bf Q_{col}}
=(0,2\pi/\sqrt{3})$. Since we are interested in computing thermodynamic quantities, the 2D spin wave theory must be extended to finite temperatures. This is not straightforward for two reasons: i) the number of excited magnons diverges within the linear spin wave theory  and, ii) the magnetization of a 2D Heisenberg model must be zero, in agreement with Mermin-Wagner's theorem\cite{mw}. In order to reconcile these two points, following Takahashi\cite{takahashi}, it is imposed a condition on the number of magnons -$S$ magnons
per site- that gives rise to a zero magnetization. Although the conventional spin wave theory relies on the existence of long range order, the elementary excitations at finite temperatures of a low dimensional antiferromagnet still resemble the spin waves excitations of the  ordered ground state. The zero magnetization
condition\cite{takahashi}   
turns out
$$ S-\frac{1}{N}\sum_k{\bf{a_k^+ a_k}}=0,
$$ 
\noindent where the $a$'s are the bosonic operators of the Holstein-Primakov spin representation, and $N$ is the number of lattice sites. Here there is one kind of boson because we have assumed that all spins are pointing in the $z$- direction of a local spin quantization axis as in ref\cite{trumper}.  Once this condition
is included by means of a Lagrange multiplier $\lambda$ in the
spin wave version of Hamiltonian(\ref{heis}),
 it can be diagonalized by means of a
Bogolyubov transformation ${\bf{\alpha_k}}=u_k{\bf{a}_k}-v_k{\bf{a}_{-k}^{+}}$.
After some algebra,
\begin{eqnarray*}
	{{H}}=\sum_k\omega_{\bf{k}}(n_{\bf{k}}+\frac{1}{2})-\lambda N (S+\frac{1}{2})+E_c(1+\frac{1}{S})
	\label{hdiagon}
\end{eqnarray*}

\noindent with a magnon energy dispersion 
$$    
 \omega_{\bf{k}}=(S/2)\sqrt{\gamma^2_{\bf{k}}-\beta^2_{\bf{k}}},
$$
 \noindent and
\begin{eqnarray*}
\gamma_{\bf{k}} &=& 2\sum_{\delta}\ J_{\delta}\ \{\cos\bf{k. \delta}\cos^2\frac{{\bf{Q.\delta}}}{2}-\cos{\bf{Q.\delta}}\}+\lambda/S \\
\beta_{\bf{k}} &=& -2\sum_{\delta}\ J_{\delta}\ \sin^2\frac{\bf{Q. \delta}}{2}\cos\bf{k\cdot \delta}. \\
\end{eqnarray*}
$E_{c} = \ N\ S^{2}\ \sum_{\delta}J_{\delta}\cos{\bf{Q.\delta}}$ is the classical energy 
 and $n_{\bf{k}}$ is the Bose occupation number. The Lagrange multiplier $\lambda$ acts as a chemical potential, opening a gap in the magnon dispersion.
 By minimizing the free energy of the system a self consistent
equation for $\lambda$ can be obtained:
\begin{eqnarray}
\frac{1}{2}\sum_k(1+2n_{\bf{k}})
({\gamma({\bf{k}}})/{\omega_{\bf{k}}})=1.
\label{Free}
\end{eqnarray}

\noindent In the modified spin wave theory the resolution of Eq. (\ref{Free}) leads  to  a temperature dependent $\lambda$, and then to a temperature dependent excitation spectrum which takes into account the effect of entropy on the system\cite{takahashi}.
In order to test the reliability of the modified spin wave theory, we have computed different finite 
temperature magnitudes like structure factor, uniform susceptibility, and specific heat.  
The static structure factor can be calculated as the Fourier transform of the mean value of the spin-spin correlation $S({\bf{q}})= \sum_{\bf{r}} e^{{\it i} {\bf{q\cdot r}}}\left\langle {\bf{S_0}}\cdot{\bf{S_r}}\right\rangle$. 
Although the modified spin wave theory is not rotationally invariant, it can be shown that the contribution of the longitudinal spin fluctuations -$S^{zz}({\bf{q}})$- reproduces exactly the  structure factor of a rotationally invariant
theory\cite{auerbach}.
The Holstein-Primakov transformation for the spin operators is replaced and a mean field decoupling is performed to terms of four boson operators. By considering only those contractions that conserve total spin\cite{takahashi}, we finally obtain,  
\begin{eqnarray*}
S({\bf{q}})=&&-\frac{1}{4}+
\frac{1}{N} \sum_{{\bf{k}}} [(u_{\bf{k+q}}^{2}+v_{\bf{k+q}}^{2})(u_{\bf{k}}^{2}+v_{\bf{k}}^{2})-\\
&&4u_{\bf{k}}\ v_{\bf{k}}\ u_{\bf{k+q}}\ v_{\bf{k+q}}\ ]
\times(n_{\bf{k+q}}+\frac{1}{2})(n_{\bf{k}}+\frac{1}{2}),
\end{eqnarray*}
while the  static uniform susceptibility is given by
\begin{eqnarray}
\chi=\frac{1}{T}\sum_{r}\left\langle S_{0}^{z}S_{r}^{z}\right\rangle=\frac{1}{3T}S({\bf{q=0}}).
\end{eqnarray}

\begin{figure}[ht]
\includegraphics[width=8.5cm]{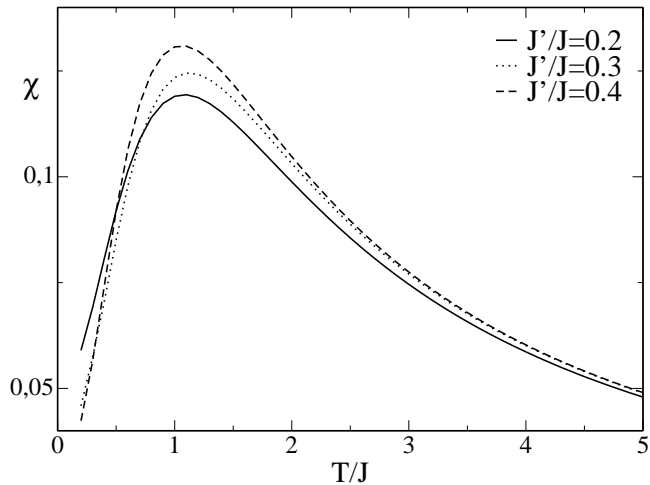}
\caption{Static uniform susceptibility versus temperature for several values of frustration.}
\label{unif}
\end{figure}
  In Fig. \ref{unif} we show the uniform susceptibility as a function of temperature for several
   values of frustration. It  has a rounded peak located at $T\sim J$ that signals the onset of the
    AF N\'eel correlations. In an actual  (3D) case this would imply AF long range order. 
    The effect of
     frustration is to slightly decrease the temperatures at which the peaks occur\cite{McKenzie-condmat}. At larger
      temperatures a Curie-like $1/T$ behavior is recovered. 
 On the other hand, as expected for a system without broken symmetry, we have corroborated that the specific heat  has also  a rounded peak and then   goes to zero as $1/T^2$ for higher temperatures. Furthermore, the structure factor
shows a  sharp peak in
the magnetic wave vector ${\bf{Q_{col}}}$ that decreases with frustration and temperature. This means that despite the
absence of long range order at finite temperature,  AF  N\'eel correlations are still dominant in our approximation.

The correct behavior obtained for the thermodynamic properties of the frustrated Heisenberg model lends support to the finite temperature approximation based on the modified spin wave theory. Regarding experiments, 
it should be noted that a quite different behavior is observed in  
the magnetic susceptibility  on $\kappa-Cl$. At low temperatures, once long range AF order is developed, a weak ferromagnetism due to the canting of the AF order has been found \cite{shimizu}. Obviously, we do not expect to capture features inherent in the three dimensionality and the anisotropy of the compound with a 2D Heisenberg model. On the other hand, for higher temperatures the Curie's law is not found experimentally. Actually, 
for temperatures higher than $40 K$ it is observed a crossover from  a paramagnetic insulator  to  a metallic phase\cite{kanoda}. In order to take into account this crossover one should include the itinerancy of the electrons in the model, but for the present study of the AF phase it is not necessary. These differences do not allow  a proper estimate of the value $J^{\prime}/J$ for the $\kappa-Cl$ compound using our model, as it was performed recently for other related compounds\cite{McKenzie-condmat}. Nevertheless, in the next section we show that the main features observed in NMR experiments  can be qualitatively well reproduced by the frustrated model.

\section{$1/T_1$ nuclear relaxation time}
The $1/T_1$  relaxation time is proportional to the transition probability of nuclear spin flips via their hyperfine interaction with the electronic spins. For an interacting localized spin model the relaxation mechanism is mediated by magnons. Single magnon processes are forbidden by energy conservation considerations but two magnon processes (Raman processes) --where one thermally excited magnon is destroyed and another is created in the nuclear spin flip process-- are allowed\cite{pincus}.  Taking into account these processes, the relaxation time $1/T_1$ can be related to the dynamic structure factor $S({\bf{q}},\omega)$ as
\begin{eqnarray}
\frac{1}{T_1}=\frac{1}{2N}\sum_{{\bf{q}},\nu} A^2_{\nu}({\bf{q}})S({\bf{q}},\omega),
\label{1/T1}
\end{eqnarray}

\noindent where $A_{\nu}({\bf{q}})$ is the hyperfine tensor and $\omega$ is Larmor nuclear frequency. In what follows we assume $A_{\nu}({\bf{q}})$ isotropic and ${\bf q}$ independent. As mentioned in section II within the linear spin wave theory $S({\bf{q}},\omega)=S^{zz}({\bf{q}},\omega)$ and  so the dynamic structure factor is defined as
 
 $$
S({\bf{q}},\omega)=\frac{1}{2 \pi}\int^{+\infty}_{-\infty}dt 
\left< \delta S^z_{\bf q}(t)  \delta S^z_{-{\bf q}}(0) \right> e^{i\omega t}.
$$
\noindent Therefore, considering only  longitudinal spin fluctuations
$$
\delta S^{z}_{\bf q}=N^{\frac{1}{2}} \sum _{{\bf q_1},{\bf q_2}} 
\delta_{{\bf q_1}+{\bf q}+{\bf Q}-{\bf q_2}} 
[a^{\dagger}_{\bf q_1} a_{\bf q_2}- \left<   a^{\dagger}_{\bf q_1} a_{\bf q_2} \right>],
$$
\noindent the two magnon structure factor results\cite{birgeneau}:

\begin{eqnarray}
	S({\bf{q}},\omega)=\sum_{\bf{k}}(u_{\bf{k}}u_{\bf{k+q}}+v_{\bf{k}}v_{\bf{k+q}})^2
	n_{\bf{k}}&&\\
	\times(1+n_{\bf{k+q}})&\delta(\omega_{\bf{k}}-\omega_{\bf{k+q}}+\omega)&.\nonumber 
\label{SF}
\end{eqnarray}
The relaxation time, Eq. (\ref{1/T1}),  can be written as \cite{rice} 

\begin{equation}
1/T_1=\frac{1}{2}\int \rho^2(\epsilon)n_{\epsilon}(1+n_\epsilon)(1+\frac{\gamma^2_{\epsilon}}{\epsilon^2})d\epsilon.
\label{DENS}
\end{equation}

\begin{figure}[h]
\includegraphics[width=3.5cm]{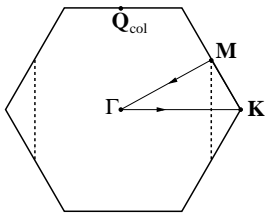}
\caption{Brillouin zone for the triangular lattice. 
The dashed lines, ${\bf k}=(\pm \pi,k_y)$, correspond to the dispersionless spin wave modes.}
\label{ZB}
\end{figure}
\begin{figure}[ht]
\includegraphics[width=8cm,angle=0]{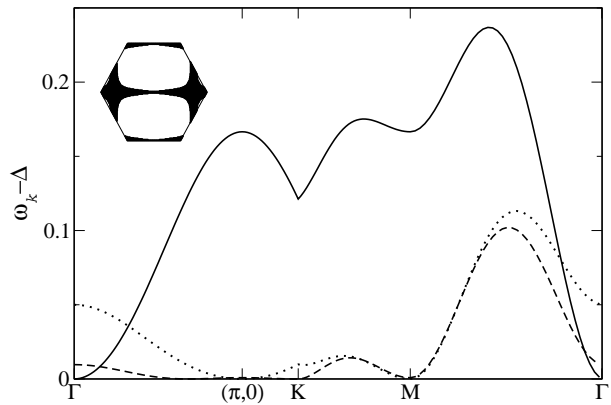}
\caption{Spin-wave dispersion, relative to the gap $\Delta$, along the $\Gamma \to K \to M \to \Gamma$ line (shown in Fig.\ref{ZB}), as a function of
temperature for $J'/J=0.2$. Solid curve: $T=3J$. Dashed curve: $T=4.4J$. Dotted curve: $T=5J$. $T=4.4J$ is the temperature of the $1/T_1$ peak.
In the inset it is shown the region of nearly degenerate spin wave modes for $T=4.4J$ (see text).}
\label{DISP}
\end{figure}


\noindent where $\rho(\epsilon)=1/N\sum_{\bf k} \delta(\epsilon-\omega_{\bf k})$ is the magnon density of states. 
For all $T$ the spin wave spectrum has two dispersionless  lines located at ${\bf k}=(\pm \pi,k_y)$ (dashed lines in Fig. \ref{ZB}) 
deriving in a van Hove singularity in the density of states. These spin wave excitations correspond to antiferromagnetic 
fluctuations along the $x$ direction. In Fig. \ref{DISP} it can be seen that as temperature increases a gap opens at the 
Goldstone modes (${\bf k}={\bf 0}, {\bf Q}_{col}$), while the energy of the dispersionless lines goes downward to the 
gap value. If the system is frustrated, at a certain temperature, the dispersionless lines reach the bottom of the spectrum, 
and a larger number of excitations becomes nearly degenerate with them (see inset of Fig. \ref{DISP}). The availability of 
this large number of nearly degenerate magnon excitations produces a sharp  increase of $1/T_1$ (the integral in eq.(\ref{DENS}) 
is dominated by the van Hove singularity of the magnon density of states). At higher temperatures the energy at 
${\bf k}={\bf 0}, {\bf Q}_{col}$ becomes greater than the energy at ${\bf k}=(\pm  \pi,k_y)$, the large degeneracy 
disappears, and $1/T_1$ goes down to a constant value inversely proportional
to $J^{\prime}/J$ . It is worth noticing that without frustration the large degeneracy mentioned above never occurs, and $1/T_1$ behaves monotonically.

\subsection{Effect of pressure}

In this section we address the role of frustration in the insulating phase of the $\kappa-Cl$, and its  possible relation with the effect of pressure.
To this end,  we compare the frustration dependence of the relaxation time predicted by the antiferromagnetic Heisenberg model 
with previous experimental results of the $\kappa-Cl$ compound under pressure\cite{lefebvre}.   
We have evaluated  eq.(\ref{1/T1}) at $\omega=0$ since for realistic values $\omega\ll J$. In the next section the model predictions  with varying $\omega$ will be discussed. 
\begin{figure}[ht]
\includegraphics[width=8cm]{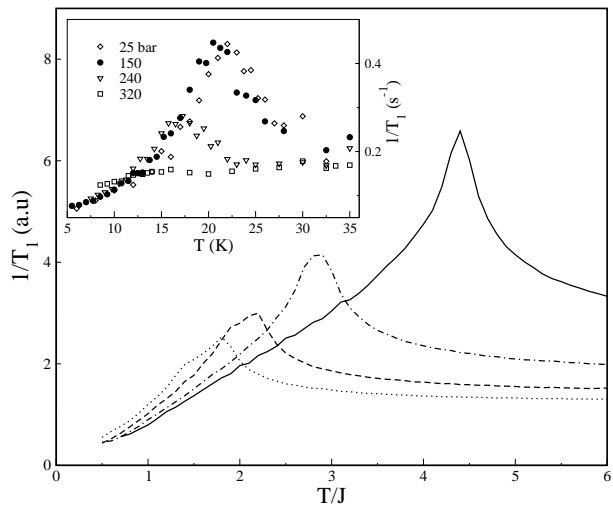}
\caption{Relaxation rate $1/T_1$ versus temperature for several values of frustration: $J^{\prime}/J=0.2$ (solid line), $0.3$ (dot-dashed), $0.4$ (dashed), and $0.48$ (dotted).
 Inset: experimental data under pressure taken from ref.\cite{lefebvre}}
\label{t1}
\end{figure}
\noindent In Fig. \ref{t1} it is shown $1/T_1$ as function of temperature for several values of frustration. 
 The effect of frustration is to suppress the $1/T_1$ peaks and move them to 
 lower temperatures. A similar behavior can be observed in previous NMR 
 experiments \cite{lefebvre} on $\kappa-Cl$  compound under pressure (see inset 
 Fig. \ref{t1}). We also find that the $1/T_1$ peak temperatures scale as 
 $J^2/J^{\prime}$. In terms of Hubbard parameters, $J=4t^2/U$, if we assume that the effect of pressure is to decrease $U/t$, then $J$ increases with pressure. Therefore, in order to reproduce the experimental behavior, $J^{\prime}/J$ should increase 
 with pressure independently of the value of $J$. This analysis  
 allows us to conjecture the existence of a close relation between the magnetic frustration and the macroscopic effect of pressure on the AF phase of $\kappa-Cl$. 
 It is worth noticing that within our model 
 it is required an appreciable variation of $J'/J$ to mimic the experimental results.
Similar conclusions has been recently pointed out by other authors in the context of Hubbard models\cite{powell05,gan05,liu05}. 
Assuming the existence of this relation and the fact that, in our theory, the $1/T_1$ peaks only appear 
with frustration, it can be said that at ambient pressure 
the $\kappa-Cl$ compound is frustrated, and that its magnetic frustration  does not change in a range of pressure up to $\sim 150 $ bar (see inset of Fig. \ref{t1}).   
 
 From Figs. \ref{unif} and \ref {t1} it can be observed that the theoretical  $1/T_1$ peaks occur at 
 higher temperatures than the uniform susceptibility ones. This is not what is expected, but we ascribe this feature to the modified spin wave approximation. In particular the constraint of zero magnetization is not rigorously implemented site by site. Instead, it has been imposed on average and this produces, at finite temperature, an underestimation of thermal fluctuations,  which is manifested in the shift to higher temperatures of the $1/T_1$ peaks. This aspect of the approximation has been also pointed out in another related frustrated model\cite{Ivanov92}.

 \subsection{Effect of magnetic field}
\begin{figure}[ht]
\includegraphics[width=8cm]{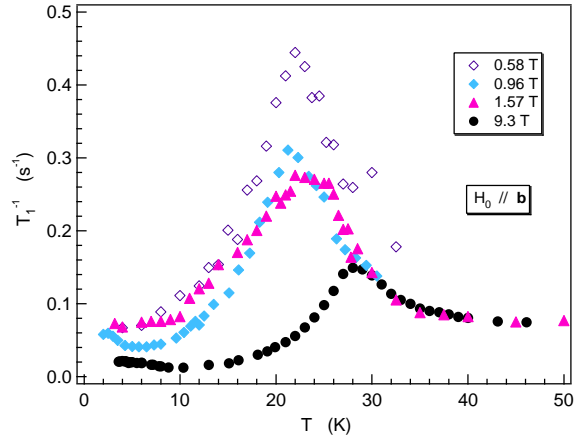}
\caption{Relaxation rate $1/T_1$  vs. magnetic field at ambient pressure.}
\label{T1H}
\end{figure} 

 \begin{figure}[ht]
\includegraphics[width=8cm]{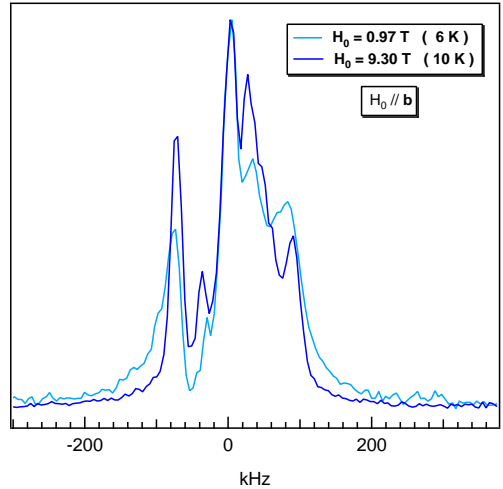}
\caption{NMR spectra of the $\kappa-Cl$ compound at ambient pressure for two different values of the magnetic field. }
\label{canting}
\end{figure}

In this section we show our measurements of the $1/T_1$ nuclear relaxation time 
under magnetic field, and we compare them with the predictions of the frustrated 
spin model. All measurements were performed on single crystals with typical dimensions
of $0.8 \times 0.7 \times 0.07$ $mm^3$.
The samples were cooled slowly ($0.35K/min$) through the $80 K$ region
 in order to avoid the effects of imperfect ethylene ordering on the ground
state\cite{Kawamoto}.  Proton spin-lattice relaxation times
have been measured using standard saturation-recovery technique where, depending on the linewidth conditions, either free induction decay or spin-echo sequence were used to record the signal intensity.\\

A feature of geometrically frustrated materials is the shift towards low energy of the spectral weight of the magnetic excitations, which could be associated with a low spin dynamics with characteristic frequencies $\omega \ll J$ \cite{Ramirez}. This behavior results in   
a strong magnetic field dependence of $1/T_1$. Indeed, it is well known that a
frequency dependent
relaxation rate is expected in systems where fluctuation timescales 
are comparable with the inverse of the Larmor frequency
\cite{Spiess}. In this section we present the $1/T_1$ frequency dependence
measurements. Figure \ref{T1H} shows the evolution of the $1/T_1$ peak for a broad range of magnetic fields. 
 It is observed a strong suppression of the $1/T_1$ peak with increasing 
 magnetic field, along with an unexpected shift towards higher temperatures of the 3D   N\'eel  temperature. 
For most of the antiferromagnetically ordered compounds the N\'eel temperature is rather stable in the range of magnetic fields we have studied, and under stronger fields  the N\'eel temperature tends to decrease. This standard evolution can be attributed to the fact that strong fields favour a more paramagnetic phase. Contrary, in our case we find an increase of the N\'eel temperature with magnetic field as well as  a rather  insensitive AF magnetization (as can be deduced from Fig. \ref{canting}). This phenomenon, which is not clearly understood, has also been observed in a related $\kappa$-compound\cite{Wzietek}.       
  
\begin{figure}[ht]
\includegraphics[width=8cm,angle=0]{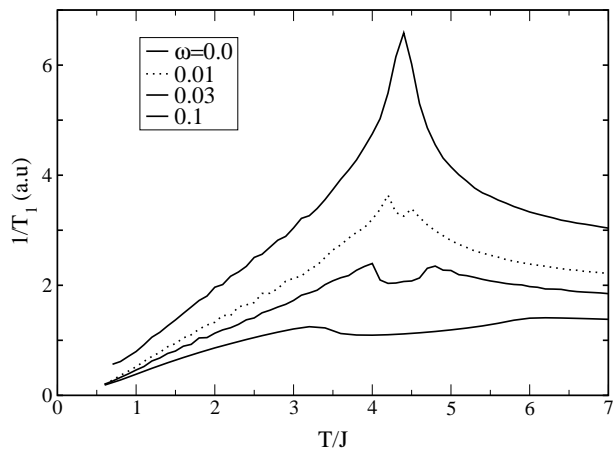}
\caption{Relaxation rate $1/T_1$ versus temperature for several values of frequency and $J^{\prime}/J=0.2$. $\omega$ is given in units of $J$.}
\label{t1w}
\end{figure}

To investigate the model prediction we have considered  the effect of the perpendicular  magnetic field  on the 
relaxation time by just varying the proton frequency $\omega$. Since the NMR spectra do not change noticeably with
magnetic field, as it is shown in  Fig. \ref{canting}, we assume in the following that the field-dependent $1/T_1$ is a {\em{frequency} 
effect}. For the same reason, we have neglected the canting of the magnetic order in the effective Hamiltonian. In  
Fig. \ref{t1w} we show the relaxation time predicted by the frustrated model for different values of frequency and a 
generic value of $J^{\prime}/J$. 
It can be seen that the frustrated model captures the strong suppression of $1/T_1$ with increasing magnetic field  observed experimentally (Fig. \ref{T1H}). The $1/T_1$ behavior is similar for other moderate frustration values, while for the unfrustrated case it does not show a strong suppression with frequency. Note that the values of $\omega$ are at least one order of magnitude lower than the magnetic frustration $J'/J$ or the gap at the peak temperature. This strong frequency dependence of $1/T_1$, characteristic of systems with slow spin dynamics, is another indication that the $\kappa-Cl$ compound at ambient pressure exhibits magnetic frustration. 

In addition, the theoretical results show a double peak structure, which becomes more pronounced for larger frequencies. As a consequence of the energy conservation of the Raman processes with $\omega \neq 0$ (i.e., the delta function in eq. (\ref{SF})) $1/T_1$ gets dominated by the large density of states around the dispersionless lines 
${\bf k}=(\pm \pi, k_y )$ and around the  ${\bf k}={\bf 0}, {\bf Q}_{col}$ modes. As temperature varies, the peaks appear when the energy difference between both regions is $\sim \pm \omega$. 
 It is worth to mention that although we capture the strong suppression of the $1/T_1$ peak, we miss  the shift of the 3D transition temperature and the magnetic field independence of  $1/T_1$ for high temperatures found experimentally (see Fig. \ref{T1H}). The latter might be due to the crossover from the paramagnetic insulator to the metallic phase above $40 K$.     
      
\section{Concluding remarks}

We have measured the magnetic field dependence of the $1/T_1$ relaxation time in the antiferromagnetic phase of the quasi-bidimensional organic compound 
$\kappa-(BEDT-TTF)_2 Cu [N(CN)_2] Cl$. We have observed a strong field dependence of $1/T_1$, characteristic of frustrated magnetic systems, due to a very slow spin dynamics\cite{Ramirez}. We have also found an unexpected shift of the N\'eel temperature towards higher temperatures under magnetic field. A similar finding has been reported 
in the $\kappa$ compound with  $X=Cu [N(CN)_2] Br$ \cite{Wzietek}.  
To analyze our NMR measurements under magnetic field, along with previous ones performed under hydrostatic pressure\cite{lefebvre} we have  used the spatially anisotropic triangular Heisenberg model as a minimal model for the AF phase of the $\kappa-Cl$. 
We have treated the spin model with a linear spin wave calculation, modified 
to take into account the basic features of the 2D antiferromagnetic spin fluctuations at finite temperature. We calculated the $1/T_1$ relaxation time considering Raman 
processes, simultaneous creation and destruction of magnons. We have studied the model 
predictions as a function of frustration and magnetic field. The latter has been considered as frequency effect, since the NMR spectra do not change noticeably in the range of the applied  magnetic fields.   
Regarding the pressure dependence of the $1/T_1$ relaxation time, we have found 
that the observed strong suppression and shift to lower temperatures of the $1/T_1$ peaks  under pressure can be qualitatively reproduced by an increase of the frustration in the Heisenberg model. Within the context of the microscopic model such shift corresponds to a genuine increase of the magnetic frustration $J'/J$, independently of the variation of $J=4t^2/U$ driven by pressure. 
These results are in line with recent works using Hubbard models, that emphasize 
the relevance of the role of frustration and its possible relation with pressure\cite{powell05,gan05,liu05}, besides the expected decrease of electronic correlation with pressure.    
On the other hand, Campos {\it et al.}\cite{campos96}, performing a H\"uckel tight binding study for a related $\kappa$ compound,  have found that the effect of hydrostatic pressure is to increase the frustration due to the  sliding  of the BEDT-TTF molecules in the dimers, one with respect to the other, along their short axes. This study also points out 
that the magnetic frustration is increased under pressure. However, the studies based on minimal microscopic models require large variation of $U/W$ and frustration with pressure for a quantitative agreement with experiments\cite{powell05}, while the electronic calculations predict a slight variation 
 of the tight binding parameters and, as a consequence, to a slight variation of 
 the correlation and frustration.\cite{mckenzie}.
A possible explanation of this discrepancy is the fact that the electronic correlation  dramatically renormalizes the bare tight binding parameters, as have been found recently by Liu {\it et al.}\cite{liu05}.
We would like to emphasize that we do not intend to give here a microscopic description of the pressure since we used an effective correlated model that do not include the actual structure of the $\kappa-Cl$ compound. The task of describing on equal foot the real structure of the compound and the electronic correlation is  a challenge outside the scope of the methods available at the present time. 
     
Concerning the magnetic field dependence of the $1/T_1$ peaks, the frustrated spin model
also reproduces the strong suppression observed experimentally for Larmor frequencies
$\omega \ll J$. This is another indication that the AF insulating phase of the $\kappa-Cl$ compound is a frustrated antiferromagnet even at ambient pressure. However we can not reproduce the shift to higher temperatures  of the $1/T_1$ peak under magnetic fields, an issue that deserves further investigation.  

We thank M. Fourmigu\'e and C. Mezi\`ere for the sample
preparation, and A. Greco and C. Bourbonnais for fruitful discussions. This work was partially supported by Fundaci\'on Antorchas.

\end{document}